\documentclass{aa}
\usepackage{savesym}
\savesymbol{bibfont}
\usepackage{amsmath}
\usepackage{lscape}
\usepackage{rotating}
\usepackage{natbib}
\usepackage{graphicx}
\usepackage{txfonts}
\setcitestyle{aysep={}}
\restoresymbol{NAT}{bibfont}

\begin{document}

\title{Gamma-Ray Burst long lasting X-ray flaring activity}

\author{
M.G. Bernardini\inst{1,2}
\and
R. Margutti\inst{1,3}
\and
G. Chincarini\inst{1,3}
\and
C. Guidorzi\inst{4}
\and
J. Mao\inst{1,5}
}

\institute{
INAF - Osservatorio Astronomico di Brera, via Bianchi 46, I-23807 Merate (LC), Italy 
\and
ICRANet, p.le della Repubblica 10, I-65100 Pescara, Italy
\and
University of Milano  Bicocca, Physics Dept., P.zza della Scienza 3, Milano 20126, Italy
\and
University of Ferrara, Physics Dept., via Saragat 1, I-44122 Ferrara, Italy
\and
Yunnan Observatory \& Key Laboratory for the Structure and Evolution of Celestial Objects, Chinese Academy of Sciences, Kunming, Yunnan Province, 650011, China 
}

\titlerunning{Late time X-ray flares}

\authorrunning{Bernardini et al.}

\date{}

\abstract
{One of the most intriguing features revealed by the \emph{Swift} satellite is the existence of flares superimposed to the Gamma-Ray Burst (GRB) X-ray light curves. The vast majority of flares occurs before $1000$ s, but some of them can be found up to $10^6$ s after the main event. }
{In this paper we shed light on \emph{late} time (i.e. with peak time $t_{pk} \gtrsim 1000$ s) flaring activity.  We address the morphology and energetic of flares in the window $\sim 10^3-10^6$ s to put constraints on the temporal evolution of the flare properties and to identify possible differences in the mechanism producing the early and late time flaring emission, if any. This requires the complete understanding of the observational biases affecting the detection of X-ray flares superimposed on a fading continuum at $t > 1000$ s. }
{We consider all the \emph{Swift} GRBs that exhibit late time flares. Our sample consists of $36$ flares, $14$ with redshift measurements. We inherit the strategy of data analysis from  Chincarini et al. (2010) in order to make a direct comparison with the early time flare properties.}
{The morphology of the flare light curve is the same for both early time and late time flares, while they differ energetically. The width of late time flares increases with time similarly to the early time flares. Simulations confirmed that the increase of the width with time is not due to the decaying statistics, at least up to $10^4$ s. The energy output of late time flares is one order of magnitude lower than the early time flare one, being $\sim 1\% E_{prompt}$.  The evolution of the peak luminosity as well as the distribution of the peak flux-to-continuum ratio for late time flares indicate that the flaring emission is decoupled from the underlying continuum, differently from early time flares/steep decay. A sizable fraction of late time flares are compatible with afterglow variability.}
{The internal shock origin seems the most promising explanation for flares. However, some differences that emerge between late and early time flares suggest that there could be no unique explanation about the nature of late time flares.}


\maketitle

\section{Introduction}\label{intro}

One of the most intriguing and unexpected features revealed by the X-Ray Telescope \citep[XRT,][]{2005SSRv..120..165B} on board the \emph{Swift} satellite \citep{2004ApJ...611.1005G} is the existence of flares superimposed to the Gamma-Ray Burst (GRB) X-ray light curves \citep{2005Sci...309.1833B,2006ApJ...641.1010F,2007ApJ...671.1903C,2007ApJ...671.1921F}. The vast majority of flares occurs before $1000$ s \citep{2007ApJ...671.1903C}, but some of them can be found up to $10^6$ s after the main event. 

Recent analyses of the flare temporal and spectral properties (\citealp{chinca10}, hereafter C10; \citealp{giantflares10}) of a large sample of \emph{early} time (i.e. with peak time $t_{pk} \lesssim 1000$ s) flares and of a subsample of bright flares revealed close similarities between them and the prompt emission pulses, pointing to an internal origin of their emission \citep{2007ApJ...671.1903C,chinca10}. Therefore, the central engine itself should remain active and variable for long time. Alternatively, flaring emission can be powered by delayed magnetic dissipation during the deceleration of the ejecta \citep{2006A&A...455L...5G}. Despite many efforts from both the observational and the theoretical point of view, a clear explanation of the X-ray flares is still missing (C10).

In this paper we shed light on \emph{late} time (i.e. $t_{pk} \gtrsim 1000$ s) flaring activity observed by XRT in the $0.3-10$ keV energy band. Its existence poses constraints on the theoretical models since late time flares require huge releases of energy ($\sim 10^{50}$ erg) up to $1$ month after the main event. A previous work on late time X-ray flares has been presented by \citet{2008A&A...487..533C}. These authors analysed a sample of $7$ GRBs that exhibit flares after $10^4$ s by fitting them with a gaussian profile plus power-law underlying continuum. The purpose was to compare the temporal and spectral properties of the underlying continuum and of the temporal and flux amplitude variability of these flares with the results found in \citet{2007ApJ...671.1903C} and with the prescription of the internal and the external shock models. They concluded that late time flares are not different from the early time ones. However, due to the small number of flares in their sample, this statement needed further investigation.

The present work addresses the morphology of flares in the temporal window $\sim 10^3-10^6$ s of a larger sample than in \citet{2008A&A...487..533C}. The good fraction of flares of the sample with redshift measurement allows us also to characterise the energetic properties at very late times. We inherit the C10 strategy of data analysis to make a direct comparison with the results obtained for early time flares and put stringent constraints on the temporal evolution of the flare properties. The aim is to identify possible differences in the mechanism producing the observed early and late time flare emission, if any.  This requires the complete understanding of the observational biases affecting the detection of X-ray flares superimposed on a fading continuum at $t > 1000$ s. 

In Sect. \ref{sample} we present the late time flare sample and we describe the fitting procedure. In Sect. \ref{analysis} we show the main results of the present analysis of the late time flares temporal profiles and we compare them with those obtained from the C10 sample. In Sect. \ref{discussion} we discuss our findings. In Sect. \ref{conclusions} we summarize the results obtained and draw our main conclusions. We adopt standard values of the cosmological parameters: $H_\circ=70$ km s$^{-1}$ Mpc$^{-1}$, $\Omega_M=0.27$ and $\Omega_{\Lambda}=0.73$.  Errors are given at $1\, \sigma$ confidence level unless otherwise stated.

\section{Sample and fitting procedure}\label{sample}

We consider all the \emph{Swift} GRBs observed between April 2005 and December 2009 that exhibit late time (i.e. $t_{pk}  \gtrsim 1000$ s) flares. Among all the X-ray light curves observed by XRT, we include in our sample only those that contain a flaring activity with a relatively complete structure: rise, peak and decay phase. The present data set consists of $22$ long GRBs (LGRB) and $1$ short GRB (SGRB), GRB050724, with $36$ flares (see Table \ref{tab1}). $9$ GRBs ($14$ flares) have measured redshift\footnote{For the redshift values we refer to  http://www.mpe.mpg.de/~jcg/grbgen.html}. These flares and their underlying continuum have been analysed in the $0.3-10$ keV energy band. For a full description of the data reduction we refer to C10.

The flare properties have been investigated by fitting the $0.3-10$ keV light curve with an empirical function proposed by \citet[][hereafter Norris05]{2005ApJ...627..324N}:
\begin{equation}
I(t)= A\, \lambda \, e^{-\frac{\tau_1}{(t-t_s)}-\frac{(t-t_s)}{\tau_2}}\, \, \, \rm{for}\, \, t\geq t_s
\end{equation}
where $\mu=(\tau_1/\tau_2)^{1/2}$ and $\lambda=e^{2\mu}$. The intensity peaks at $t_{pk}=\tau_{pk}+t_s=(\tau_1 \tau_2)^{1/2}+t_s$. The pulse width is measured between the two $1/e$ intensity points,
\begin{equation}
w=\Delta t_{1/e}=t_d+t_r=\tau_2(1+4\,\mu)^{1/2}\, .
\end{equation}
The pulse asymmetry is
\begin{equation}
k=\frac{t_d-t_r}{t_d+t_r}=(1+4\,\mu)^{-1/2}\, .
\end{equation}
The decay time $t_d$ and the rise time $t_r$ are expressed in terms of $w$ and $k$ as:
\begin{equation}
t_{d,r}=\frac{1}{2}w(1\pm k)\,.
\end{equation}
The choice of the Norris05 profile allows us to perform a direct comparison of the properties of late and early time flares.

We fitted simultaneously the underlying continuum and the flares by adopting a multiply broken power-law plus an appropriate number of Norris05 functions. As a measure of the quality of the fit we used the $\chi^2$ statistics. The results of the fitting procedure are available as online material. The errors on the derived quantities are obtained accounting for the entire covariance matrix of the fitting parameters. The flare fluence $S$ is calculated by integrating the corresponding Norris05 function over the entire duration. The count -to -flux conversion factor has been computed from a spectrum extracted during the entire duration of the burst \citep{2007A&A...469..379E,2009MNRAS.397.1177E,2010MNRAS.402...46M}.

\section{Analysis and results}\label{analysis}

In the following we characterise the late time flare emission by investigating the flare morphology and energetics, that the theoretical models most commonly address. The morphology of the flare light curves is presented in the observer frame, while the energetic properties are investigated in the source rest frame.

\subsection{Morphology}\label{lc}

\begin{figure}
\includegraphics[width=\hsize,clip]{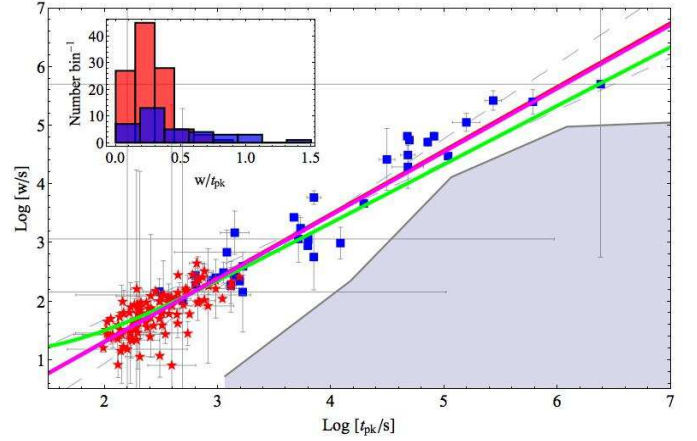}
\caption{Width vs. peak time for the sample of late time flares (blue squares) compared with early time flare sample (red stars). Green line: linear ($w=10.+0.2\,t_{pk}$) behaviour found in the early flare sample. Pink line: best-fitting for the late flares sample, $w=10^{(-0.9\pm0.7)}\,t_{pk}^{(1.1\pm0.2)}$. Gray area: region of the $w-t_{pk}$ plane that cannot be explored in our analysis. \emph{Inset:} distribution of the ratio $w/t_{pk}$ of late time flares (blue rectangles) compared with the early time flare one (red rectangles). The two distributions are centered on a similar value (the median is $0.28$ for the late time flares and $0.23$ for the early time flares) but late time flares have a larger standard deviation ($\sigma=0.33$ compared with $\sigma=0.14$).}
\label{widthtpk}
\end{figure}

Flares are erratic emission of huge amount of energy on a relatively short timescale. The median of the distribution of the ratio between the width $w$ and the time $t_{pk}$ at which flares occur is $<w/t_{pk}>=0.28$, that is similar to the one of the population of early time flares ($0.23$). However, late time flares have a much broader distribution ($\sigma=0.33$ compared with $\sigma=0.14$). The highest $w/t_{pk}$ occurs for the flare belonging to the only SGRB of our sample (GRB050724).

In Fig. \ref{widthtpk} we show the evolution of $w$ with time for late time flares. The width depends on time as $w\propto t_{pk}^{1.1}$. The increase of the flare duration with time observed in this sample as well as in the early time flare sample is one of the main differences from the prompt emission pulses: while $w$ is almost constant with time for the BATSE pulses belonging to the Norris05 sample, it linearly increases for the early time flare sample (see Fig. 3 in C10). 

An important issue is if this increasing trend is a real feature or it is due to the progressive widening of the light curve temporal bins: as the time proceeds, the decaying light curve is built by integrating the observed counts over larger time intervals in order to keep the signal to noise ratio constant. Therefore, an intense and narrow flare would be spread into a less intense and wider one, thus resulting in a possible bias for the correlation. Simulations in \citet{2007ApJ...671.1903C} revealed indeed that very short duration flares (up to a factor $10^{-2}$ with respect to the best fitting relation shown in Fig. \ref{widthtpk}) could be in principle detected at $10^5$ s at $90\%$ confidence level. This result has been obtained by generating flares with any possible luminosity, width and temporal occurrence. The real situation is slightly different, since flares become dimmer as the time proceeds (see Sect. \ref{lum}). This motivated us to perform simulations that reproduce more in detail the observed scenario. 

We simulate for each flare of the sample a set of fake flares by adding to the best fitting of the continuum different Norris05 profiles that have the same peak intensity and peak time of the original flare but with different decreasing widths. Then, we rebin each simulated light curve as the corresponding observed one, and we fit the results with a new Norris05 profile plus the underlying continuum. We define in this way a minimum observable width for each time interval as the width calculated with the best fitting Norris05 parameters of the shortest simulated flare occurring in that time interval that we can identify after the rebinning procedure. Low significance fluctuations (number of $\sigma$ $\lesssim 2$) have not been classified as flares. The result is shown in Fig. \ref{widthtpk}: the gray area represents the forbidden region for flare detection, namely a flare with those width and peak time values would be classified as a non significant fluctuation around the underlying continuum. We found that the $w-t_{pk}$ best fitting relation is unbiased up to $10^4$ s since the forbidden region is at least $1$ order of magnitude below the best fitting relation, consistently with \citet{2007ApJ...671.1903C}. After $\sim 10^4$ s the range of detectable widths reduces progressively, being only $24\%$ of the observed value at $10^6$ s. This is due to the progressively lower statistics of the GRB afterglows with time, which possibly causes a departure from the best fitting $w-t_{pk}$ relation found at early times.

Another source of bias affecting in some cases the late time flare sample is  the difficulty to distinguish blended structures. In fact the integration of the observed counts over progressively larger time intervals makes harder and harder to discriminate if the observed excess is due to one single flare or it comes from the contribution of many overlapping flares. Therefore in some cases the width obtained with our fitting procedure could be overestimated but only by a factor of a few. This factor will be not so different from the average number of flares observed in a temporal interval $\Delta t\sim t$ of the early time sample, that is $\sim 1.5$.

\begin{figure}
\includegraphics[width=\hsize,clip]{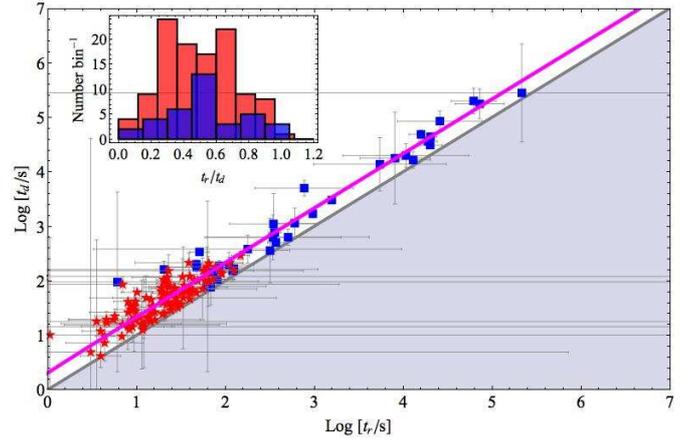}
\caption{Decay time vs. rise time for the sample of late time flares (blue squares) compared with the early time flare sample (red stars). The best linear fitting for the two samples is $t_d=(-0.1\pm0.6)+(2.15\pm0.04)\, t_r$ (pink line). Gray area: $t_d\leqslant t_r$. \emph{Inset:} distribution of the ratio $t_r/t_d$ of late time flares (blue rectangles) compared with the early time flare one (red rectangles). The two distributions are centered on a similar value (median $0.52$ with $\sigma=0.23$ for the late time flares and median $0.49$ with $\sigma=0.26$ for the early time flares).}
\label{risedec}
\end{figure}

A remarkable property of the flare profile is a \emph{self-similar} behaviour, being the rise over decay time constant in time. This feature is found in the present sample of late time flares as well as in the C10 analysis, with the same best fitting relation  $t_d\sim 2\,t_r$ and a distribution of $t_r/t_d$ (see Fig. \ref{risedec}) and of asymmetry $k$ (median value $0.31$ with standard deviation $0.22$) with a similar central tendency and dispersion. The K-S test on the two distributions of $t_r/t_d$ for early time and late time flares gives a $60\%$ probability that they belong to the same population. Still, there is no flare having $t_r>t_d$ (gray area in Fig. \ref{risedec}).

\subsection{Energy and luminosity}\label{lum}

\begin{figure}
\includegraphics[width=\hsize,clip]{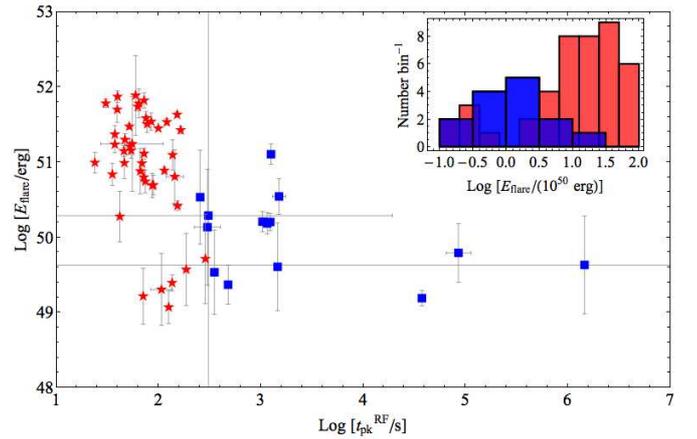}
\caption{Isotropic energy emitted in the observer $0.3-10$ keV energy band vs. rest frame peak time for the sample of late time flares (blue squares) compared with the early time flare sample (red stars). The absence of energetic late time flares is manifest. \emph{Inset:} distribution of the logarithm of the isotropic energy emitted in the $0.3-10$ keV energy band of late time flares (blue rectangles) compared with the early time flare one (red rectangles). The late time flares are clearly less energetic than early time ones, with a median for the distribution of the energy $1.4\times 10^{50}$ erg and standard deviation $\sigma=3.2\times 10^{50}$ erg.}
\label{energy}
\end{figure}

\begin{figure}
\includegraphics[width=\hsize,clip]{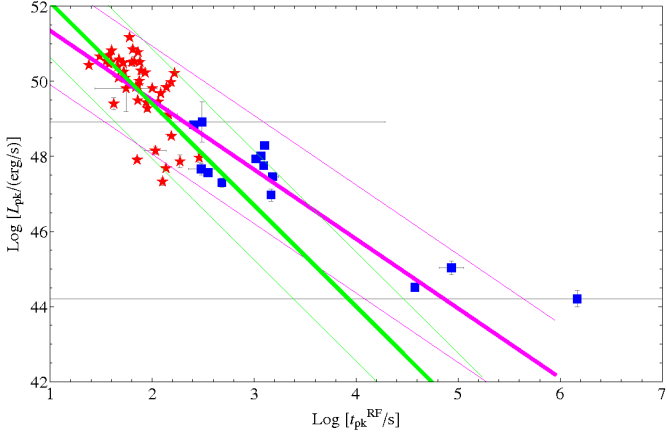}
\caption{Peak luminosity vs. peak time measured in the rest frame for the $14$ late time flares with measured redshift (blue squares) compared with the early time flare sample (red stars). The best fitting for the whole sample accounting for the extrinsic scatter is $Log[L_{pk}]=(53.2\pm0.3)+ (-1.8\pm0.1)\,Log[t_{pk}]$ with $\sigma_{ext}=(0.72\pm0.07)$ (pink line). The best fitting for early time flares accounting for the extrinsic scatter $Log[L_{pk}]= (54.8\pm4.0)+(-2.7\pm0.5)\,Log[t_{pk}]$ with $\sigma_{ext}=(0.73\pm0.08)$ (green line) is shown for comparison. The thinner lines indicate the $2\sigma$ regions for the two best fitting relations.}
\label{lumtpk}
\end{figure}

In Fig. \ref{energy}, \emph{Inset} the distribution of the logarithm of the isotropic energy emitted in the observer frame $0.3-10$ keV energy band of late time flares is shown, compared with that obtained in C10 for early time flares. Late flares are less energetic, with a median value of $1.4\times 10^{50}$ erg, one order of magnitude lower than the early time flare sample. As stated in C10, the distribution function of the isotropic energy emitted is likely incomplete at the low-energy tail. On the other hand it is evident that high energetic flares are not detected at late times (see Fig. \ref{energy}). 

One of the most important clue on the understanding of X-ray flares is the evolution of the peak luminosity $L_{pk}$ versus time. In Fig. \ref{lumtpk} we compare the behaviour $L_{pk}\propto t_{pk}^{-2.7}$ found in C10 with our sample: the flares with $t_{pk}^{RF}>10^4$ s differ from this relation for more than $2\sigma$. The best fitting of the late time flare luminosity-time relation leads to $L_{pk}\propto t_{pk}^{-1.2}$. This result is consistent with the average luminosity behaviour after $1000$ s found in \citet[][hereafter M10]{lummedia}. In M10 is clearly shown that at this epoch the average luminosity function of flares tracks the detectability threshold. Therefore, the unbiassed luminosity-time relation for late flares is likely steeper. If we perform a joined fitting of the early and late time flare data, we find $L_{pk}\propto t_{pk}^{-1.8}$. If we account for an extra-variance of the sample  \citep{2005physics..11182D}, this best-fitting relation has a $\chi^2/dof=58.1/55$ and a p-value of $40\%$. 

\begin{figure}
\includegraphics[width=\hsize,clip]{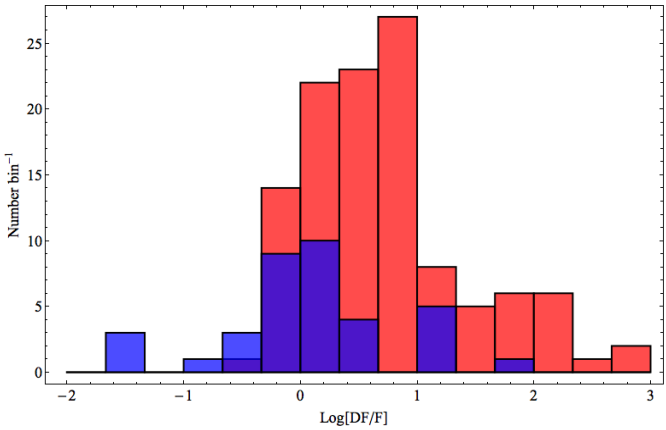}
\caption{Distribution of the logarithm of the peak flux to continuum ratio of late time X-ray flares (blue rectangles) compared with the early time flare one (red rectangles). The late time flares have clearly a smaller $\Delta F/F$ than early time ones, with a median for the distribution of the logarithm $0.04$ compared with $0.60$ for the early time flares, and standard deviation $\sigma=0.75$ compared with $\sigma=0.73$.}
\label{fluxvar}
\end{figure}

In M10 we reported that the distribution of the peak flux-to-continuum ratio $\Delta F/F$ of the C10 early time flare sample can be fitted by the superposition of two gaussian profiles. If we include the late time flares of the present sample, we obtain that they all lie in the "fainter" part of the distribution (see Fig. \ref{fluxvar}). A K-S test performed on the two distributions gives a $2\times 10^{-5}$ probability that they belong to the same population. 

The smaller ratio at late times can be ascribed to two occurrences: a) the peak flux of flares decreases with time, then at late times flares are fainter (see also Fig. \ref{lumtpk}); b) after $1000$ s in $80\%$ of the light curves of our sample the underlying continuum changes its behaviour becoming shallower. To investigate these two effects, we calculate how the $\Delta F/F$ of a late time flare would appear at early times. The ratio $\Delta F/F$ is substituted for the equivalent peak luminosity to continuum ratio $\Delta L/L$. We assume that the peak luminosity of flare evolves as $t^{-2.7}$ for $t\leqslant1000$ s and $t^{-\alpha}$ for $t>1000$ s, while the continuum evolves as $t^{-2.7}$ for $t\leqslant1000$ s (M10) and $t^{-1}$ for $t>1000$ s, this index being the average behaviour between $1000$ s and $10^4$ s of the light curves of our sample. We leave $\alpha$ as a free parameter due to the impossibility to evaluate quantitatively the luminosity-time relation for late time flares. However, it is reasonably between the value of the detection threshold found in M10 and the value found for early time flares, $1.2< \alpha \leqslant 2.7$. We consider a flare with $\Delta L/L = 1$, equal to the median value for the late flare sample distribution, at $10^4$ s. Extrapolating backwards, we find that, depending on $\alpha$, the corresponding $\Delta L/L$ at $10^2$ s would be larger up to $50$, within the range of values observed for the early time sample. This simple calculation reveals that the discrepancy in the $\Delta L/L$ values for late time and early time flares depends on the shallower behaviour of the continuum after $1000$ s that results in a bias for the flare detection (M10).

\section{Discussion}\label{discussion}

The existence of flares at $10^6$ s after the trigger poses several constraints on GRB central engines. Many authors pointed out that late time flares must require late ejection of shells rather than late collisions of prompt emission shells \citep{2006ApJ...642..354Z,2007MNRAS.375L..46L,2009ApJ...707.1623M}. This conclusion, based on the constraints imposed by the observed duration of flares \citep{2007MNRAS.375L..46L}, implies that the collisions producing flares occur in the same conditions as those producing the prompt emission pulses, but the ejection time is comparable to the flare observational time  \citep{2007MNRAS.375L..46L}. Therefore, the differences observed in flares with respect to the prompt emission pulses depend entirely on intrinsic differences in the shells ejected by the central engine and not on their evolution \citep{2009ApJ...707.1623M}. More important, the central engine itself should remain active up to $10^6$ s. 

Late time flares are less energetic than early flares by at least $1$ order of magnitude (see Fig. \ref{energy}, \emph{Inset}). In Fig. \ref{energy} it is manifest that early time flares from C10 span a wide range of energies, and a hint of bimodality appears in their energy distribution (see C10, Fig. 8, \emph{Inset}). Late time flares belong mainly to the low energy tail of the distribution. This translates, within the internal shock model, into less energetic collisions at late times \citep{2008MNRAS.388L..15L,2009ApJ...707.1623M}, and that corresponds to less energetic shells ejected at late times from the central engine \citep{2008MNRAS.388L..15L}. 

The peak luminosity of early time flares decreases with time as a power-law $L_{pk}\propto t_{pk}^{-2.7} $, as found in C10 (see Fig. \ref{lumtpk}) and obtained also in M10 with a different method. The late time flares of the present sample observed at $10^3\lesssim t\lesssim 10^4$ s are slightly brighter with respect to the expectations based on the early time flares, but still consistent within the dispersion (see Fig. \ref{lumtpk}). This is mainly a result of the bias introduced by the underlying continuum: at $t\sim 1000$ s we observe in $90\%$ of the X-ray light curves of our subsample of GRBs that show late time flares and with a redshift measurement the transition from a steep decay to a shallow decay phase \citep{2009MNRAS.397.1177E}, and this prevents the flares belonging to the faint tail of the distribution from being detected.

A different situation occurs for the three flares\footnote{Note however that one of the $3$ flares with $t_{pk}\gtrsim 10^5$ s belongs to the short GRB050724.} observed at $t\gtrsim 10^5$ s that are well above the extrapolation of the $t^{-2.7}$ behaviour. To understand the nature of these flares we estimate the probability that they are compatible with the distribution of the observed early time flares around the extrapolated fitting $L_{pk}\propto t^{-2.7}$ with the same extra-variance of the sample $\sigma_{ext} = 0.73$.  We compute the logarithmic average X-ray luminosity light curve of the underlying continuum of the GRBs of our sample, and we define the minimum luminosity for a flare at a given time as the one that can be detected at $2 \sigma$ over the average continuum light curve. At $t=10^5$ s the probability that a flare belonging to that distribution has a luminosity greater than the minimum one is less than $10^{-5}$. At $t=10^6$ s this probability is even much lower, $<10^{-9}$. 

Conversely, we could assume that all the flares of the C10 and of the present sample follow the same luminosity-time relation, but it is not the one found in C10 (and, we recall, also in M10 with a different method). In this case for the joined sample of early time and late time flares it reads: $Log[L]=53.2-1.8\,Log[t]$, with an extrinsic scatter $\sigma_{ext}=0.72$. This is an acceptable fit ($\chi^2/dof=58.1/55$, p-value$=40\%$). However, in M10 we showed that the early time flare average luminosity  between $60$ s and $400$ s is steeper than $t^{-1.8}$ at $90\%$ confidence level. Moreover the continuum after $1000$ s introduces a severe bias for the detectability of faint late time flares: probably only the brightest flares can overshine the continuum at late times (M10). Therefore, the best fitting $t^{-1.8}$ likely overestimates the peak luminosity of flares. 

If we consider the distribution of the peak flux to continuum ratio $\Delta F/F$ for late time and early time flares, we find that the two datasets differ significantly, with a K-S probability that they belong to the same population of $2\times 10^{-5}$. The faintness of the $\Delta F/F$ for late time flares can be ascribed to the shallower behaviour of the underlying continuum after $1000$ s: with a simple calculation we show that the same flare with $\Delta F/F\sim 1$ at $10^4$ s would have a $\Delta F/F\sim 50$ at $10^2$ s due to the steeper continuum. This under the crucial assumption that the evolution of the peak luminosity of late time flares is decoupled from the underlying continuum, differently from what has been found for early time flares. In M10 we found that the average luminosity of flares at $t\lesssim 400$ s decreases with time with a power-law index that is linearly correlated to the one of the underlying steep decay. This points to a deep connection between the mechanisms that produce the two components. If this connection is not established at late times, it is natural to expect that late time flares are so rare, since only the brightest ones are able to overshine the continuum (M10).

In the framework of accretion models, the long-lasting accretion up to late times can be due either to a very small viscosity parameter $\alpha\lesssim 0.01$, or to a continued mass fall-back \citep{2008MNRAS.388.1729K}. The first possibility seems to be disfavored since it predicts temporal indices for the initial and final decays of the whole light curve less steep than the observed ones \citep{2008MNRAS.388.1729K}. In the second scenario, in a very short timescale a quasi steady-state is established with $\dot{m}_{acc}\sim \dot{m}_{fb}$, where $\dot{m}_{acc}$ and $\dot{m}_{fb}$ are the accretion rate and the mass fall-back rate respectively; therefore the observed luminosity tracks the mass fall-back rate fairly well \citep{2008MNRAS.388.1729K}. The fall-back radius is $r_{fb}>1000\,r_s$, where $r_s$ is the corresponding Schwarzschild radius (for a central black hole mass $M_{BH}=5 M_{\odot}$), and an accretion timescale of $\sim 10^4$ s is not unreasonable for very low accretion rates \citep{2006ApJ...636L..29P}, that is the case of \citet{2008MNRAS.388.1729K}, Fig. 7. 

In this context, the presence of flares in the light curve is due to sudden changes in the accretion rate \citep{2008MNRAS.388.1729K}. The duration of each accretion episode, that corresponds to the duration of the flare, is of the order of the accretion time \citep{2006ApJ...636L..29P} with $t_{acc}/t$ often less than unity \citep[see][Fig. 6-7]{2008MNRAS.388.1729K}. Disk instabilities could in principle lead to large variations in the accretion rate. A promising possibility is gravitation instabilities (we refer to M10 for a discussion on thermal and viscous instabilities in the context of early time flares): these instabilities may occur in the outer regions of the disk ($r>100\,r_s$) if the angular momentum is high. This requirement is more likely fulfilled in collapsars than in the merging of compact objects \citep{2006ApJ...636L..29P}. Alternatively, the fall-back material can fragment into bound objects producing a modulation in the accretion rate (\citealp{2008MNRAS.388.1729K}; see \citealp{2007MNRAS.376L..48R} in the context of the merging of compact objects). If fragments form, their initial mass depends strongly on the local properties. However, their estimated mass is $M_{frag}\sim 10^{50}$ erg for fragmentation taking place at very large radii/times \citep[][eq. 5]{2006ApJ...636L..29P}, that corresponds to the median energy observed for late time flares (see Fig. \ref{energy}, \emph{Inset}).

A different process of fragmentation can occur in the collapse of a rapidly rotating stellar core, with the formation of more than one compact object \citep{2005ApJ...630L.113K}. These subsequently coalesce under the effect of gravitational radiation releasing an energy $\sim 2\%$ of the main event with a delay of hundreds of seconds \citep{2002ApJ...579L..63D,2005ApJ...630L.113K}. The reactivated burst has a jet axis close to the first one, thus it expands in a cleaner environment with lower baryonic contamination and consequently with an higher Lorentz factor \citep{2005ApJ...630L.113K}. This can explain why flares are softer than the main event. 

Many GRBs have X-ray light curves without flares. In fact a flaring activity is observed in $\sim 33\%$ of XRT light curves. The conditions for fragmentation in the outer layer of the disk are uncertain, however high rotation is generally required \citep{2006ApJ...636L..29P}. Rotation plays an important role in shaping the X-ray light curve at very early stages \citep{2008MNRAS.388.1729K}: Fig. 5 of \citet{2008MNRAS.388.1729K} shows clearly that lowering the rotation rate of the central object, the resulting light curve exhibits a longer and fainter prompt duration with a steeper decline. However the rotation rate cannot be very low (less than $8$ times lower than the model in Fig. 2 of \citealp{2008MNRAS.388.1729K}) otherwise the entire core directly collapses to a black hole preventing the occurrence of the GRB. We noted in M10 that GRBs with multiple flares have generally a much flatter steep decay light curve. A flatter decay can be produced by a higher rotation rate for these GRB central objects, that allows to produce more suitable conditions for fragmentation.

For late time X-ray light curve the angular momentum of the fall-back material is crucial to shape the transition between the shallow decay phase and the following normal decay. If the angular momentum is low, this transition results in a sudden drop of the light curve. In the opposite case the transition is smooth. In our sample of GRBs with late time flares, only $3$ cases show a final decay steeper than $1.5$. This is consistent with the expectations of high angular momentum for the fall-back material, and to the leading role that it has in favoring the fragmentation. A complete analysis of the light curves without flares needed in order to confirm this scenario is beyond the scope of the present work and will be presented elsewhere.

Magnetic fields strongly modify the dynamics of accretion \citep[see e.g.][]{2003ApJ...592..767P}. In particular the energy release can be repeatedly stopped and restarted by the magnetic flux accumulated around the central object, giving origin to the observed X-ray flares, if the mass supply rate decreases with time \citep{2003ApJ...592..767P,2006MNRAS.370L..61P}. In this context a flaring episode depends on the strength of the magnetic field, namely when the magnetic stress provides support against gravity: $2B_R B_Z/4\pi \sim GM\Sigma/r^2$, where $B$ is the magnetic field, $M$ the mass of the central object, $r$ the radius and $\Sigma$ the surface density. A magnetic flux of the order of $\phi=\pi r^2 B(r)\sim 10^{29}$ cm$^2$ G is necessary to stop accretion with rate $10^{-3}$ M$_\odot$ s$^{-1}$ around a $3$ M$_\odot$ central object at $r=300\,r_s$, representative of the late time evolution \citep{2006MNRAS.370L..61P}. Simple estimates show for the emission episodes that they are less energetic of one or two orders of magnitude than the prompt emission, and the duration of each episode is coupled with the time of occurrence \citep{2006MNRAS.370L..61P}, thus reproducing the main observational features of flares.

Despite the similarity with the early time flares and the prompt emission pulses points to an internal shock origin of late time flares, thus favoring accretion models, there are different promising possibilities as the magnetic reconnection model \citep{2006MNRAS.369L...5L,2006A&A...455L...5G,2009ApJ...695L..10L} that do not require a long lasting activity of the central engine. In fact in this model the emission takes place just before the deceleration radius. Within this context, flares are originated by delayed magnetic dissipations during the deceleration of the ejecta \citep{2006A&A...455L...5G}. For a constant variability (which is indeed the case of flares) the isotropic energy emitted is proportional to the energy in the forward shock \citep{2006A&A...455L...5G}. This is consistent with the fact that late time flares are on average less energetic than early time ones. However, the time dependence of the width can be recovered only with ad-hoc assumptions on the characteristic length scales of the reconnection regions \citep{2006A&A...455L...5G,2009ApJ...695L..10L}.

The linear evolution of the decay time as a function of the rise time is present in the early as well as in the late time flare sample, being $t_d\sim 2\,t_r$ (see Fig. \ref{risedec}). This is a result that is valid over $5$ orders of magnitude, and it is also property common to the prompt emission pulses as shown in C10. This is a very stringent requirement for any theoretical model, since the rise and decay times depend either on the dynamic of the shocks within the internal shock model \citep{1997ApJ...490...92K}, or on the motion of turbulent eddies or mini-jets \citep{2009MNRAS.394L.117N,2009ApJ...695L..10L}.

\subsection{Flares associated to variability in the GRB afterglow: the case of GRB050724}

\begin{figure}
\includegraphics[width=\hsize,clip]{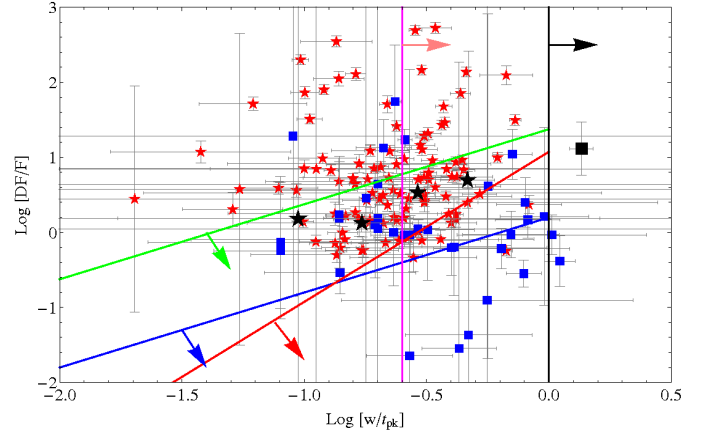}
\caption{Increase of the flux with respect to the underlying continuum vs. temporal variability for the sample of late flares (blue squares) compared with the early flare sample (red stars). The black stars and square correspond to the SGRB early time and late time flares, respectively. The solid lines limit the kinematically allowed regions for density fluctuations on-axis (blue line) and off-axis (red line), for off-axis multiple regions density fluctuations (green line), for patchy shell model (black line) and for refreshed shocks (pink line), see \citet{2005ApJ...631..429I} for details.}
\label{Ioka}
\end{figure}

In the above discussion we focused mainly on an internal origin of flares. However, it is of some interest to investigate late time flares in the context of variability of the GRB afterglow. A useful tool in this respect is the plot proposed by \citet{2005ApJ...631..429I} where they trace the regions of allowance for the increase of the flux with respect to the continuum as a function of the temporal variability for bumps in the afterglow on the basis of kinematic arguments. In Fig. \ref{Ioka} we present our sample of late time flares in a "Ioka plot", compared with the results presented in Fig. 13 of C10. As stated in Sect. \ref{lc} the range of observed $w/t_{pk}$ of the two sample is comparable, while as we noted in Fig. \ref{fluxvar} the flux increase with respect to the continuum covers a different range in the case of late time flares. 

All the late time flares of our sample, with the exception of that of GRB050724, fulfill the general requirement for internal shocks $\Delta t/t<1$. However, $\sim 14\%$ are compatible with internal shock origin \emph{only}. Differently from the early time flares, $\sim 25\%$ of late time flares are compatible also with bumps due to on-axis density fluctuations. This reveals that, although the internal shock origin is the most promising explanation for flares, a sizable fraction of late time flares can have a different origin than early time ones. Therefore, there could be no unique explanation about the nature of late time flares. Similar results were found by \citet{2008A&A...487..533C} that compared the flare-to-continuum flux versus variability of the $8$ flares of their sample\footnote{We recall that the width in the Norris05 profile is measured at $\sim 37\%$ of the maximum. If we consider the corresponding width measured on a Gaussian profile, it is $w_{gauss}=2.83\sigma$, while the Full Width at Half Maximum (FWHM) is $w_{FWHM}=2.35\sigma$. Therefore, the position on the x axis of the flares in the Ioka plot crucially depends on the profile adopted and on the definition of the duration. In the work of \citet{2008A&A...487..533C} the duration of flares is the FWHM.} with the sample of \citet{2007ApJ...671.1903C}. They found that one flare (i.e. $\sim 12\%$ of their sample) can be explained only with internal shocks, and all the other flares can be interpreted both by internal and by external shocks. However, they do not have any flare compatible with on-axis density fluctuations.

The flare of the short GRB050724 is the most challenging. As pointed out in Sect. \ref{lc} it has the longest duration among the flares of the joined early and late sample, $w/t_{pk}\sim 1.4$. This translates in the Ioka plot in being the only case compatible with the geometrical constraint for the multicomponent jet model, the refreshed shock model and marginally with the density fluctuation of many regions.

GRB050724 has also been observed at the epoch of the flare in the optical and radio frequencies \citep[][and references therein]{2007A&A...473...77M}. The simultaneous analysis of X-ray and optical data led \citet{2007A&A...473...77M} to conclude that they likely belong to the same spectral component (with $\beta_{OX}=(0.72\pm0.04)$) that implies a similar temporal behaviour. Therefore, despite the poor sampling of the optical light curve, they identify the presence of a simultaneous flare in the optical energy band \citep{2007A&A...473...77M}. A rebrightening in the radio light curve is also present, but it is delayed with respect to the X-ray and optical peaks \citep{2007A&A...473...77M}. 

The association of this flaring emission in different energy bands with the afterglow has been tested by \citet{2006MNRAS.367L..42P}. He included the flare in the forward shock emission in an interstellar medium environment as an episode of energy injection and simulated the observed X-ray, optical and radio light curves. He found agreement with the observations, however using different assumptions and a smaller dataset than in \citet{2007A&A...473...77M}. In the same context of the external shock model, the observed duration of the flare and spectral index can be accommodated in the adiabatic shock propagating in a wind environment \citep[][Fig. 2]{2007MNRAS.375L..46L}. 

The simultaneous detection of a flare at different wavelengths is not a feature common to all the X-ray flares (see e.g. C10). Moreover, \citet{giantflares10} found in two cases of very bright flares that during the emission episode, the peak energy crosses the X-ray energy band with a temporal dependence very close to the temporal decay of the observed flux. Although this has not been demonstrated to be a common property of all flares, the optical-X-ray SED reported by \citet{2007A&A...473...77M} seems to exclude that the peak frequency crosses the X-ray band. 

The above reasoning reveals indeed that the flare observed in the X-ray light curve of the short GRB050724 can be interpreted within different frameworks. Also the early time flares associated to SGRBs included in the C10 sample are all compatible with afterglow variability (see Fig. \ref{Ioka}). While consensus has been reached on the origin of flares associated to long GRBs, there is no overwhelming evidence on the nature of the flares in SGRBs. A devoted study is necessary in order to increase our knowledge on the SGRB progenitors and their environment.

\section{Summary and conclusions}\label{conclusions}

We analysed $23$ GRB X-ray light curves observed by XRT that exhibit late time (i.e. $t_{pk}  \gtrsim 1000$ s) flares. The present work inherits the C10 strategy of data analysis but extends the temporal window of investigation of three orders of magnitude, from $10^3$ s to $10^6$ s, with a larger sample than in previous works on late time flares \citep{2008A&A...487..533C}. We found that late time flares are similar to early time flares for the following reasons:

\begin{itemize}
\item the width of late time flares increases with time similarly to the early time flares. This is a real property of flares, at least up to $10^4$ s;
\item the \emph{self-similar} behaviour of the flare profile, being the decay proportional to the rise time, is also a property shared by both early time and late time flares. This is a more stringent requirement than being simply "asymmetrical pulses".
\end{itemize}
However, they differ for the following characteristics:
\begin{itemize}
\item the median energy output of late time flares is one order of magnitude lower than the early time flare one, being $\sim 1\% E_{prompt}$;
\item the bias introduced by the shallower underlying continuum after $1000$ s allows the detection of only the brightest flares. However it seems unlikely that the flares observed at $t\gtrsim 10^5$ s are compatible with the extrapolation of the behaviour $t^{-2.7}$ found for early time flares;
\item the decoupling of the evolution of the peak luminosity of late time flares from the underlying continuum can account for the fainter distribution of the peak flux to continuum ratio $\Delta F/F$ of late time flares;
\item a sizable fraction of late time flares are compatible with variability in the GRB afterglow.
\end{itemize}

The global pictures that emerges is that the morphology of the flare light curve is the same for both early and late time flares, while they differ energetically. The similarities of late time flares with the early time ones and with the prompt emission pulses as well as their energetic can be fairly well explained in the framework of the accretion models. The existence of flares in this context is ascribed to instabilities either in the disk  \citep{2006ApJ...636L..29P} or in the fall-back material \citep{2008MNRAS.388.1729K,2007MNRAS.376L..48R}. This fragmentation can in principle account for the observed energetic and the longer duration of the accretion episodes at later times \citep{2006ApJ...636L..29P}. Intermittence can be achieved also in presence of magnetic fields \citep{2003ApJ...592..767P,2006MNRAS.370L..61P}. Different explanations are provided by the magnetic reconnection model \citep{2006MNRAS.369L...5L,2006A&A...455L...5G,2009ApJ...695L..10L} that do not require a long lasting activity of the central engine. However, the presence of $\sim 86\%$ of the late time flares of our sample that can be ascribed to variability in the GRB afterglow emission leads us to conclude that there could be no unique explanation about the nature of late time flares.

\section*{Acknowledgments}
We thank the referee J. Norris for valuable comments and suggestions. MGB thanks P. D'Avanzo, E. Zaninoni and A. Melandri for useful discussions.
This work is supported by ASI grant SWIFT I/011/07/0, by  the
Ministry of University and Research of Italy (PRIN MIUR 2007TNYZXL), by
MAE and by the University of Milano Bicocca (Italy).
This work made use of data supplied by the UK Swift Science Data Centre at the University of Leicester.


\begin{thebibliography}{34}
\expandafter\ifx\csname natexlab\endcsname\relax\def\natexlab#1{#1}\fi

\bibitem[{{Burrows} {et~al.}(2005{\natexlab{a}}){Burrows}, {Hill}, {Nousek},
  {Kennea}, {Wells}, {Osborne}, {Abbey}, {Beardmore}, {Mukerjee}, {Short},
  {Chincarini}, {Campana}, {Citterio}, {Moretti}, {Pagani}, {Tagliaferri},
  {Giommi}, {Capalbi}, {Tamburelli}, {Angelini}, {Cusumano}, {Br{\"a}uninger},
  {Burkert}, \& {Hartner}}]{2005SSRv..120..165B}
{Burrows}, D.~N., {Hill}, J.~E., {Nousek}, J.~A., {et~al.} 2005{\natexlab{a}},
  Space Science Reviews, 120, 165

\bibitem[{{Burrows} {et~al.}(2005{\natexlab{b}}){Burrows}, {Romano}, {Falcone},
  {Kobayashi}, {Zhang}, {Moretti}, {O'Brien}, {Goad}, {Campana}, {Page},
  {Angelini}, {Barthelmy}, {Beardmore}, {Capalbi}, {Chincarini}, {Cummings},
  {Cusumano}, {Fox}, {Giommi}, {Hill}, {Kennea}, {Krimm}, {Mangano},
  {Marshall}, {M{\'e}sz{\'a}ros}, {Morris}, {Nousek}, {Osborne}, {Pagani},
  {Perri}, {Tagliaferri}, {Wells}, {Woosley}, \&
  {Gehrels}}]{2005Sci...309.1833B}
{Burrows}, D.~N., {Romano}, P., {Falcone}, A., {et~al.} 2005{\natexlab{b}},
  Science, 309, 1833

\bibitem[{{Chincarini} {et~al.}(2010){Chincarini}, {Mao}, {Margutti},
  {Bernardini}, {Guidorzi}, {Pasotti}, {Giannios}, {Valle}, {Moretti},
  {Romano}, {D'Avanzo}, {Cusumano}, \& {Giommi}}]{chinca10}
{Chincarini}, G., {Mao}, J., {Margutti}, R., {et~al.} 2010, MNRAS, 406, 2113

\bibitem[{{Chincarini} {et~al.}(2007){Chincarini}, {Moretti}, {Romano},
  {Falcone}, {Morris}, {Racusin}, {Campana}, {Covino}, {Guidorzi},
  {Tagliaferri}, {Burrows}, {Pagani}, {Stroh}, {Grupe}, {Capalbi}, {Cusumano},
  {Gehrels}, {Giommi}, {La Parola}, \& {Mangano}}]{2007ApJ...671.1903C}
{Chincarini}, G., {Moretti}, A., {Romano}, P., {et~al.} 2007, ApJ, 671, 1903

\bibitem[{{Curran} {et~al.}(2008){Curran}, {Starling}, {O'Brien}, {Godet}, {van
  der Horst}, \& {Wijers}}]{2008A&A...487..533C}
{Curran}, P.~A., {Starling}, R.~L.~C., {O'Brien}, P.~T., {et~al.} 2008, A\&A,
  487, 533

\bibitem[{{D'Agostini}(2005)}]{2005physics..11182D}
{D'Agostini}, G. 2005, arXiv:physics/0511182

\bibitem[{{Davies} {et~al.}(2002){Davies}, {King}, {Rosswog}, \&
  {Wynn}}]{2002ApJ...579L..63D}
{Davies}, M.~B., {King}, A., {Rosswog}, S., \& {Wynn}, G. 2002, ApJ, 579, L63

\bibitem[{{Evans} {et~al.}(2009){Evans}, {Beardmore}, {Page}, {Osborne},
  {O'Brien}, {Willingale}, {Starling}, {Burrows}, {Godet}, {Vetere}, {Racusin},
  {Goad}, {Wiersema}, {Angelini}, {Capalbi}, {Chincarini}, {Gehrels}, {Kennea},
  {Margutti}, {Morris}, {Mountford}, {Pagani}, {Perri}, {Romano}, \&
  {Tanvir}}]{2009MNRAS.397.1177E}
{Evans}, P.~A., {Beardmore}, A.~P., {Page}, K.~L., {et~al.} 2009, MNRAS, 397,
  1177

\bibitem[{{Evans} {et~al.}(2007){Evans}, {Beardmore}, {Page}, {Tyler},
  {Osborne}, {Goad}, {O'Brien}, {Vetere}, {Racusin}, {Morris}, {Burrows},
  {Capalbi}, {Perri}, {Gehrels}, \& {Romano}}]{2007A&A...469..379E}
{Evans}, P.~A., {Beardmore}, A.~P., {Page}, K.~L., {et~al.} 2007, A\&A, 469,
  379

\bibitem[{{Falcone} {et~al.}(2006){Falcone}, {Burrows}, {Lazzati}, {Campana},
  {Kobayashi}, {Zhang}, {M{\'e}sz{\'a}ros}, {Page}, {Kennea}, {Romano},
  {Pagani}, {Angelini}, {Beardmore}, {Capalbi}, {Chincarini}, {Cusumano},
  {Giommi}, {Goad}, {Godet}, {Grupe}, {Hill}, {La Parola}, {Mangano},
  {Moretti}, {Nousek}, {O'Brien}, {Osborne}, {Perri}, {Tagliaferri}, {Wells},
  \& {Gehrels}}]{2006ApJ...641.1010F}
{Falcone}, A.~D., {Burrows}, D.~N., {Lazzati}, D., {et~al.} 2006, ApJ, 641,
  1010

\bibitem[{{Falcone} {et~al.}(2007){Falcone}, {Morris}, {Racusin}, {Chincarini},
  {Moretti}, {Romano}, {Burrows}, {Pagani}, {Stroh}, {Grupe}, {Campana},
  {Covino}, {Tagliaferri}, {Willingale}, \& {Gehrels}}]{2007ApJ...671.1921F}
{Falcone}, A.~D., {Morris}, D., {Racusin}, J., {et~al.} 2007, ApJ, 671, 1921

\bibitem[{{Gehrels} {et~al.}(2004){Gehrels}, {Chincarini}, {Giommi}, {Mason},
  {Nousek}, {Wells}, {White}, {Barthelmy}, {Burrows}, {Cominsky}, {Hurley},
  {Marshall}, {M{\'e}sz{\'a}ros}, {Roming}, {Angelini}, {Barbier}, {Belloni},
  {Campana}, {Caraveo}, {Chester}, {Citterio}, {Cline}, {Cropper}, {Cummings},
  {Dean}, {Feigelson}, {Fenimore}, {Frail}, {Fruchter}, {Garmire}, {Gendreau},
  {Ghisellini}, {Greiner}, {Hill}, {Hunsberger}, {Krimm}, {Kulkarni}, {Kumar},
  {Lebrun}, {Lloyd-Ronning}, {Markwardt}, {Mattson}, {Mushotzky}, {Norris},
  {Osborne}, {Paczynski}, {Palmer}, {Park}, {Parsons}, {Paul}, {Rees},
  {Reynolds}, {Rhoads}, {Sasseen}, {Schaefer}, {Short}, {Smale}, {Smith},
  {Stella}, {Tagliaferri}, {Takahashi}, {Tashiro}, {Townsley}, {Tueller},
  {Turner}, {Vietri}, {Voges}, {Ward}, {Willingale}, {Zerbi}, \&
  {Zhang}}]{2004ApJ...611.1005G}
{Gehrels}, N., {Chincarini}, G., {Giommi}, P., {et~al.} 2004, ApJ, 611, 1005

\bibitem[{{Giannios}(2006)}]{2006A&A...455L...5G}
{Giannios}, D. 2006, A\&A, 455, L5

\bibitem[{{Ioka} {et~al.}(2005){Ioka}, {Kobayashi}, \&
  {Zhang}}]{2005ApJ...631..429I}
{Ioka}, K., {Kobayashi}, S., \& {Zhang}, B. 2005, ApJ, 631, 429

\bibitem[{{King} {et~al.}(2005){King}, {O'Brien}, {Goad}, {Osborne}, {Olsson},
  \& {Page}}]{2005ApJ...630L.113K}
{King}, A., {O'Brien}, P.~T., {Goad}, M.~R., {et~al.} 2005, ApJ, 630, L113

\bibitem[{{Kobayashi} {et~al.}(1997){Kobayashi}, {Piran}, \&
  {Sari}}]{1997ApJ...490...92K}
{Kobayashi}, S., {Piran}, T., \& {Sari}, R. 1997, ApJ, 490, 92

\bibitem[{{Kumar} {et~al.}(2008){Kumar}, {Narayan}, \&
  {Johnson}}]{2008MNRAS.388.1729K}
{Kumar}, P., {Narayan}, R., \& {Johnson}, J.~L. 2008, MNRAS, 388, 1729

\bibitem[{{Lazar} {et~al.}(2009){Lazar}, {Nakar}, \&
  {Piran}}]{2009ApJ...695L..10L}
{Lazar}, A., {Nakar}, E., \& {Piran}, T. 2009, ApJ, 695, L10

\bibitem[{{Lazzati} \& {Perna}(2007)}]{2007MNRAS.375L..46L}
{Lazzati}, D. \& {Perna}, R. 2007, MNRAS, 375, L46

\bibitem[{{Lazzati} {et~al.}(2008){Lazzati}, {Perna}, \&
  {Begelman}}]{2008MNRAS.388L..15L}
{Lazzati}, D., {Perna}, R., \& {Begelman}, M.~C. 2008, MNRAS, 388, L15

\bibitem[{{Lyutikov}(2006)}]{2006MNRAS.369L...5L}
{Lyutikov}, M. 2006, MNRAS, 369, L5

\bibitem[{{Malesani} {et~al.}(2007){Malesani}, {Covino}, {D'Avanzo}, {D'Elia},
  {Fugazza}, {Piranomonte}, {Ballo}, {Campana}, {Stella}, {Tagliaferri},
  {Antonelli}, {Chincarini}, {Della Valle}, {Goldoni}, {Guidorzi}, {Israel},
  {Lazzati}, {Melandri}, {Pellizza}, {Romano}, {Stratta}, \&
  {Vergani}}]{2007A&A...473...77M}
{Malesani}, D., {Covino}, S., {D'Avanzo}, P., {et~al.} 2007, A\&A, 473, 77

\bibitem[{{Margutti} {et~al.}(2010{\natexlab{a}}){Margutti}, {Bernardini},
  {Barniol Duran}, {Guidorzi}, {Shen}, \& {Chincarini}}]{lummedia}
{Margutti}, R., {Bernardini}, G., {Barniol Duran}, R., {et~al.}
  2010{\natexlab{a}}, MNRAS, accepted (arXiv:1009.0172)
  
\bibitem[{{Margutti} {et~al.}(2010{\natexlab{b}}){Margutti}, {Genet}, {Granot},
  {Duran}, {Guidorzi}, {Chincarini}, {Mao}, {Schady}, {Sakamoto}, {Miller},
  {Olofsson}, {Bloom}, {Evans}, {Fynbo}, {Malesani}, {Moretti}, {Pasotti},
  {Starr}, {Burrows}, {Barthelmy}, {Roming}, \&
  {Gehrels}}]{2010MNRAS.402...46M}
{Margutti}, R., {Genet}, F., {Granot}, J., {et~al.} 2010{\natexlab{b}}, MNRAS,
  402, 46

\bibitem[{{Margutti} {et~al.}(2010{\natexlab{c}}){Margutti}, {Guidorzi},
  {Chincarini}, {Bernardini}, {Genet}, {Mao}, \& {Pasotti}}]{giantflares10}
{Margutti}, R., {Guidorzi}, C., {Chincarini}, G., {et~al.} 2010{\natexlab{c}},
  MNRAS, 406, 2149  

\bibitem[{{Maxham} \& {Zhang}(2009)}]{2009ApJ...707.1623M}
{Maxham}, A. \& {Zhang}, B. 2009, ApJ, 707, 1623

\bibitem[{{Narayan} \& {Kumar}(2009)}]{2009MNRAS.394L.117N}
{Narayan}, R. \& {Kumar}, P. 2009, MNRAS, 394, L117

\bibitem[{{Norris} {et~al.}(2005){Norris}, {Bonnell}, {Kazanas}, {Scargle},
  {Hakkila}, \& {Giblin}}]{2005ApJ...627..324N}
{Norris}, J.~P., {Bonnell}, J.~T., {Kazanas}, D., {et~al.} 2005, ApJ, 627, 324

\bibitem[{{Panaitescu}(2006)}]{2006MNRAS.367L..42P}
{Panaitescu}, A. 2006, MNRAS, 367, L42

\bibitem[{{Perna} {et~al.}(2006){Perna}, {Armitage}, \&
  {Zhang}}]{2006ApJ...636L..29P}
{Perna}, R., {Armitage}, P.~J., \& {Zhang}, B. 2006, ApJ, 636, L29

\bibitem[{{Proga} \& {Begelman}(2003)}]{2003ApJ...592..767P}
{Proga}, D. \& {Begelman}, M.~C. 2003, ApJ, 592, 767

\bibitem[{{Proga} \& {Zhang}(2006)}]{2006MNRAS.370L..61P}
{Proga}, D. \& {Zhang}, B. 2006, MNRAS, 370, L61

\bibitem[{{Rosswog}(2007)}]{2007MNRAS.376L..48R}
{Rosswog}, S. 2007, MNRAS, 376, L48

\bibitem[{{Zhang} {et~al.}(2006){Zhang}, {Fan}, {Dyks}, {Kobayashi},
  {M{\'e}sz{\'a}ros}, {Burrows}, {Nousek}, \& {Gehrels}}]{2006ApJ...642..354Z}
{Zhang}, B., {Fan}, Y.~Z., {Dyks}, J., {et~al.} 2006, ApJ, 642, 354

\end{thebibliography}
\end{document}